\documentclass[dvips]{article}

\usepackage{icrc2011}

\title{Particle acceleration and the origin of gamma-ray emission from Fermi
Bubbles.\\ {\it (Expanded version of the talk at the 32nd ICRC)}}

\newcommand{\etal}{\MakeLowercase{\textit{et al. }}} % "et al."
\shorttitle{D.Chernyshov \etal Particle acceleration in Fermi bubbles}

\authors{D.O. Chernyshov$^{1,2,3}$, K.-S. Cheng$^{2}$, V.A. Dogiel$^{1}$, C.M. Ko$^{2,3}$, W.-H. Ip$^{3}$ and Y. Wang$^2$}
\afiliations{$^1$I.E.Tamm Theoretical Physics Division of P.N. Lebedev Institute of Physics, Leninskii pr. 53, 119991, Moscow, Russia\\ $^2$Department of Physics, The University of Hong Kong, Pokfulam Road, Hong Kong, China\\$^3$Institute of Astronomy, National Central University, Jhongli 320, Taiwan}
\email{chernyshov@td.lpi.ru}

\abstract{Fermi LAT has discovered two extended gamma-ray
bubbles above and below the galactic plane.  We propose that
their origin is due to the energy release in the Galactic
center (GC) as a result of quasi-periodic star accretion onto the central black
hole. Shocks generated by these processes propagate into the
Galactic halo and accelerate particles there.  We show that
  electrons accelerated up to $\sim 10$ TeV may be responsible for the observed
gamma-ray emission of the bubbles as a result of inverse Compton (IC)
scattering on the relic photons.  We also suggest that the Bubble could
generate the flux of CR protons at energies $> 10^{15}$ eV because
the shocks in the Bubble have much larger length scales and longer
lifetimes in comparison with those in SNRs. This may explain the the
CR spectrum above the knee.}


\begin{document}
\maketitle

\section{Introduction}
Fermi bubbles are symmetric structures elongated above and below
the Galactic plane for about 8 kpc \cite{meng}. Their discovery is
one of the most remarkable events in astrophysics. The origin of
the bubble is still enigmatic and up to now a few models were
presented in the literature. The team, which subtracted this
structured gamma-ray emission from the total diffuse galactic
emission presented different explanations of the phenomenon though
they seem to trend towards the model of a single huge energy
release in the Galactic center (GC) when about 10 million years
ago a huge cloud of gas or a star cluster was captured by the
central black hole that produced the Fermi bubbles seen today.
Gamma-rays in their model are produced by the inverse Compton
scattering of the electrons on relic photons. In \cite{merch} it
was assumed that these electrons were accelerated in the bubble
interior by a hypothetic MHD turbulence excited behind the
termination shock generated at the edges of the bubbles. Another
interpretation was presented by \cite{crock} who proposed a
relatively slow energy release ($\sim 10^{39}$ erg s$^{-1}$) due
to supernova explosions as a source of proton production in the GC
which emitted gamma-rays there.

An alternative explanation based on the assumption that the energy
for the Fermi bubbles was originated from star capture events which
occurred in the GC every $10^4-10^5$ years was suggested by
\cite{cheng}. About one thousands of such events form giant
shocks propagating through the central part of the Galactic halo
and thus produce accelerated particle responsible for the bubble
emission.

Processes of particle acceleration by the bubble shocks in terms
of sizes of the envelope, maximum energy of accelerated particles,
etc. may differ significantly from those obtained for SNs that may
lead to the maximum energy of accelerated particles much larger
than can be reached in SNRs. In this respect, we assume that
acceleration of protons in Fermi bubbles may contribute to the
total flux of the Galactic cosmic rays (CR) above the ``knee'' break
($\geq 10^{15}$eV).

\section{Structure of Shocks in the Fermi Bubble}\label{struct}
As it was assumed in \cite{cheng} the central massive black hole
captures a star every $\tau_0\sim 10^4-10^5$ years and as a result
releases energy about $\mathcal{E}_0\sim 10^{52}$ erg in the form
of subrelativistic particles which heat the central $\leq 100$ pc
in the GC. This heating produces a shock propagating into the
surrounding medium \cite{weav77}. For an exponential atmosphere
with the plasma density profile
\begin{equation}
\rho(z)=\rho_0\exp\left(-\frac{z}{z_0}\right)\,,
\end{equation}
an analytical solution for permanent energy injection by star
accretion was obtained by \cite{komp}. Here $z$ is the coordinate
perpendicular to the Galactic plane. The radius of the bubble as a
function of the height $z$ and the time $t$ is
\begin{equation}
r=2z_0\arccos\left[\frac{1}{2}e^\frac{z}{2z_0}\left(1-\left(\frac{y}{2z_0}\right)^2+
e^{-\frac{z}{z_0}}\right)\right]\,,\label{r_shock}
\end{equation}
where
\begin{equation}
y=\int\limits^t_0\left(\frac{\gamma_g^2-1}{2}\lambda\frac{\alpha
Wt}{V(t)\rho_0}\right)^{0.5}dt\,,
\end{equation}
$V$ is a current volume enveloped by the shock
\begin{equation}
V(t)=2\pi\int\limits^{a(t)}_0r^2(z,t)dz\,,
\end{equation}
$a$ is the position of the shock top
\begin{equation}
a(t)=-2z_0\ln\left(1-\frac{y}{2z_0}\right)\,,
\end{equation}
$W=\mathcal{E}_0/\tau_0$ is the average luminosity of the central
source, $\gamma_g$ is the polytropic coefficient, and $\alpha$ and
$\lambda$ are numbers \cite{kogan}.

\begin{figure}
\centering
\includegraphics[width=2.in]{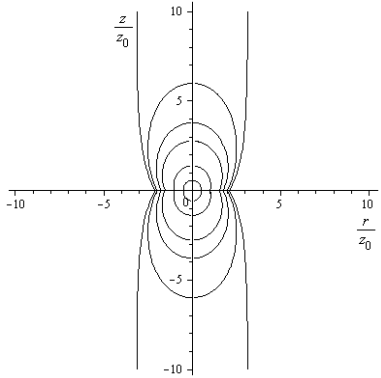} \caption{ The bubble
multi-shock structure.} \label{shape}
\end{figure}

Depending on capture parameters one can imagine the bubble
interior as a volume filled with many shocks of different ages
(see Fig.\ref{shape}).

\section{Particle Acceleration by the Bubble Shocks}
Correct analysis of shock acceleration in the bubble requires
sophisticated calculations of each stage of this process which we
hope to perform later. Now we present simple estimates of
characteristics of accelerated spectra. We analyze  spectra of
protons and electrons separately  because they are formed by
completely different processes and as we later conclude their
spectral characteristics differ strongly from each other. We
present also the spectrum of gamma-ray emission produced by the
inverse Compton (IC) scattering of the accelerated electrons in
the bubble.

\subsection{Proton Spectrum}
Below we analyse the spectrum of protons accelerated in the bubble
and discuss whether the bubble contribution to the total flux of
CRs in the Galaxy may explain the knee steepening. We remind that
the generally accepted point of view is that the flux of
relatively low energy CRs ($<10^{15}$eV) is generated by SNRs
which ejects a power-law spectrum $E^{-2}$ into the interstellar
medium. This spectrum is steepened by propagation (escape)
processes in the Galaxy in accordance with the spectrum observed
near Earth (for details see \cite{ber90}). However these sources
can hardly produce CRs with energies more than $10^{15}$eV
(see \cite{bell,volk}). Just at this energy a steepening (the
knee) in the CR spectrum is observed. We assume that the bubble
may generate the flux of CRs at energies above $10^{15}$eV.
The origin of CRs with energies above $10^{15}$eV is the goal of
this subsection. It is natural to assume that the bubble shocks
may produce this flux considering their large scales and long
lifetimes.

In the framework of our model one can imagine the bubble as at
cylinder of the radius $\sim 3-6$ kpc filled with hundreds of
shocks propagating in series one after another (see Fig.
\ref{shape}). The frequency of shock injection (star capture) is
poorly known parameter, especially for the conditions of the GC.
From the analysis of shock hydrodynamics we estimated the age of
outer shock by $10^8$ yr though this value depends strongly on
parameters of the self-similar solution which is an idealization of conditions in the Galaxy.  The frequency of star
capture by a central black hole is estimated from theoretical
treatments $\nu\sim 10^{-4}-10^{-5}$yr$^{-1}$ \cite{alex05}. Thus,
a multi-shock structure can be produced by the capture processes
with an average distance $L$ between separate shocks given by
\begin{equation}
L=\tau_{cap}u=30(\nu\times 3\times 10^4yr)(u/10^8cm/s)pc.
\end{equation}

On the other hand there is another spatial scale which
characterizes processes of particle acceleration by a single shock
which is $l_D\sim D/u$ where $D$ is the spatial diffusion
coefficient near a shock and $u$ is the  shock velocity.
Depending on the relation between  $L$ and $l_D$ there may be
different regimes of acceleration in the bubble.  This problem of
particle acceleration in supersonic turbulence or
multi-shock structure was analyzed in  \cite{byk90}. The
acceleration regime is characterized by a dimensionless
parameter
\begin{equation}
\psi=\frac{L}{l_D}=\frac{uL}{D}\,,
\end{equation}
If $\psi<<1$,  the diffusion length scale of a single shock $l_D$
exceeds the distance between shocks $L$. Therefore, particle are
accelerated by interactions with many shocks that gives the
acceleration similar to the classical stochastic Fermi
acceleration.  The critical energy $E_1$ for this regime of
acceleration is estimated from the condition $\psi\sim 1$ or
$l(E_1)\sim L$. In the Bohm limit with the diffusion coefficient
$D\sim c r_L(E)/3$ where $r_L(E)=E/eB$ is the particle Larmor, the
value of $E_1$ is
\begin{equation}
E_1\sim \frac{eBLu}{c} = 10^{15}\left(\frac{B}{5\mu
G}\right)\left(\frac{L}{30pc}\right)\left(\frac{u}{10^8cm/s}\right)\rm
eV. \label{e1}
\end{equation}
which is about the position of the knee break.

The kinetic equation for the case $\psi<<1$ ($E>>E_1$) only contains
the diffusion in the spatial and momentum spaces (Fermi
stochastic acceleration) and has the form
\begin{eqnarray}
&&\frac{\partial}{\partial z}\left(D(\rho,p)\frac{\partial
f}{\partial z}\right) +\frac{1}{\rho}\frac{\partial}{\partial
\rho} \left(D(\rho,p)\rho\frac{\partial
f}{\partial\rho}\right)+\nonumber\\
&&+\frac{1}{p^2}\frac{\partial}{\partial p}\left(\kappa(\rho,p)
p^2\frac{\partial f}{\partial p}\right)=-Q(\rho,z,p)\,,
\label{dif_bubble}
\end{eqnarray}
where $\rho$ and $z$ are the cylindrical spatial coordinates, $p$
is the particle momentum. $D(\rho,p)$ is the spatial diffusion
coefficient and $\kappa(\rho,p)$ is the momentum diffusion
coefficient,
\begin{equation}
\kappa\sim \frac{u^2}{cL}p^2 \,. \label{kappa}
\end{equation}
The term $Q(\rho,z,p)$ describes  CR injection by supernova
remnants in the disk with energies $E<10^{15}$ eV and with observed power-law
spectrum.

\begin{figure}
\centering
\includegraphics[width=3.in]{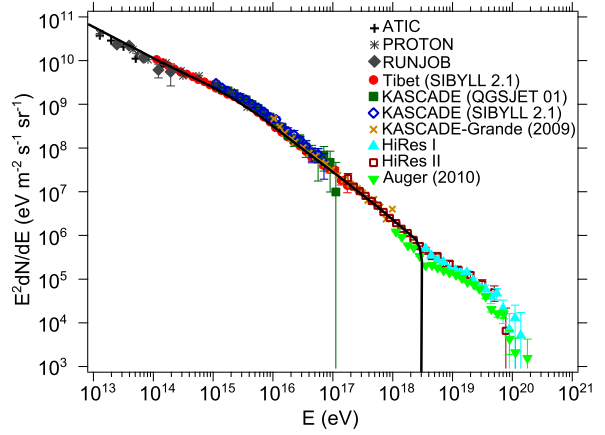} \caption{The Bubble contribution to the flux of CRs. Tibet, KASCADE, HiRes and Auger data are summarized in \cite{kumiko}. ATIC data are from \cite{ahn08}, Proton data are from \cite{gri71}, RUNJOB data are from \cite{apa01}.} \label{fig:cr}
\end{figure}

The momentum dependence of $f$ can be presented by a power-law
function, $f(p)\propto p^{-\gamma}$, where $\gamma$ should be
determined from Eq.(\ref{dif_bubble}):
\begin{equation}
\gamma\simeq\frac{3}{2}+\sqrt{\frac{9}{4}+\frac{cLD_0}{u^2H^2}}
\end{equation}
where $H$ is the height of the Galactic halo and $D = D_0
(E/E_0)^\xi$ is the average diffusion coefficient in the Galaxy.
For $H\simeq 10$ kpc, $D_0\simeq 3\times 10^{29}$ cm$^2$s$^{-1}$,
$\xi\simeq 0.3$, $E_0\simeq 3$ GeV \cite{strong} and $u = 10^8$
cm/s, $L = 30$ pc we have $\gamma\simeq 5$, i.e., the power-law
spectrum above $10^{15}$ eV is in the form $N(E)\propto E^{-3}$ as
necessary for the knee spectrum. The  proton spectrum derived from
Eq. (\ref{dif_bubble}) is shown in Fig. \ref{fig:cr}.

\subsection{Electron Spectrum and Spatial Distribution}
Unlike protons relativistic electrons lose their
energy effectively by synchrotron and inverse Compton radiation. The  rate of
energy loss of electrons is
\begin{equation}
\frac{dE}{dt}=-\beta
E^2=-c\sigma_T\left(w_H+w_{ph}\right)\left(\frac{E}{m_ec}\right)^2\mbox{.}
\label{el_sp}
\end{equation}
The maximum energy of electrons $E_{max}$ can be estimated if the
spatial diffusion coefficient at the shock $D_{sh}$ is known. From
Eq. (\ref{el_sp}) we have
\begin{equation}\label{eemax}
E^e_{max}\sim \frac{u^2c}{\beta D_{sh}} \label{Emax}
\end{equation}
For the Bohm diffusion coefficient we can estimate the upper limit
for $E^e_{max}$ which is
\begin{equation}
E^e_{Bmax}\sim\sqrt{\frac{eHu^2}{3c\beta}}\,.
\end{equation}
that gives for the velocity $u=10^8$ cm s$^{-1}$, the magnetic
field strength $H=10^{-5}$ G and $w_{ph}=0.25$ eV cm$^{-3}$ the
maximum energy of accelerated electrons about $E_{Bmax}\sim
5\times 10^{13}$eV, i.e., for electrons $E<E_{Bmax}<<E_1$. It
follows from this inequality that electrons are accelerated in the
regime $\psi>>1$ (single shock acceleration). Kinetic equations
for this case were derived in \cite{byk90}.

In the case of electrons there is one more spatial parameter
essential for acceleration by shocks, namely the electron mean
free path $\lambda$. Therefore we have two more dimensionless
parameters which describe electron acceleration in the case of
supersonic turbulence, $l_{sh}/\lambda$ and $l_D/\lambda$. Nearby
shocks electrons propagate by  diffusion, and $\lambda$ there is a
function of  energy
\begin{equation}
\lambda_D(E)\sim\sqrt{\frac{D}{\beta E}} \label{lambda_D}
\end{equation}
From Eq. (\ref{Emax}) one can see that for $E<E^e_{max}$ we have
$\lambda
> l_D$, i.e., acceleration by shocks is effective.

Far away from shocks electrons propagate by convection and for the
velocity $u$ the mean free path is
\begin{equation}
\lambda_V(E)\sim\frac{u}{\beta E} \label{lambda_V}
\end{equation}
 For relatively low
energy electrons $\lambda(E)>l_{sh}$, and  a regime of
 multi-shock acceleration for electrons is realized in this
 case where interactions with many shocks change slowly the shock to $E^{-1}$
  (for details see \cite{byk90})
   which produces  a hard
  $E^{-1}$ spectrum . The break position $E^\ast$
between the single shock and multi-shock  spectra  follows from
the equality  $\lambda(E^\ast)=l_{sh}$
\begin{equation}
E^\ast\sim \frac{u}{\beta l_{sh}}
\end{equation}
that gives, e.g., $E^\ast\sim 2.8\times 10^{11}$ eV for
$l_{sh}=10^{21}$ cm.

\begin{figure}[ht]
\centering
\includegraphics[width=3.in]{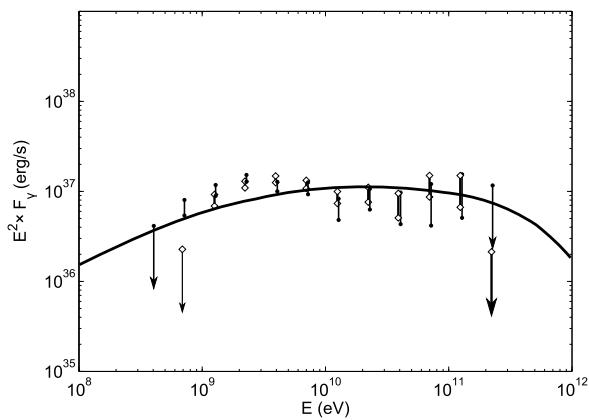} \caption{The spectrum of
gamma-ray emission from Fermi bubble in case of multi-shock acceleration. Data
points are taken from \cite{meng}.} \label{fig:gamma}
\end{figure}

 If we assume that the bubble gamma-rays are produced by IC
scattering of electrons on the relic photons, then it follows from
our model that: 1. a drop of the bubble gamma-ray flux at
$E_\gamma<1$ GeV as observed by \cite{meng} is due to a flattening
of electron spectrum at $E<E^\ast$; and 2. a drop of this spectrum
in the range $E_\gamma>100$ GeV is due to a cut-off in the
electron spectrum at $E\sim E^e_{max}$. The expected spectrum of
gamma-ray emission from the Bubble due to IC scattering of the
electrons is shown in Fig. \ref{fig:gamma}.

\begin{figure}[ht]
\centering
\includegraphics[width=3.in]{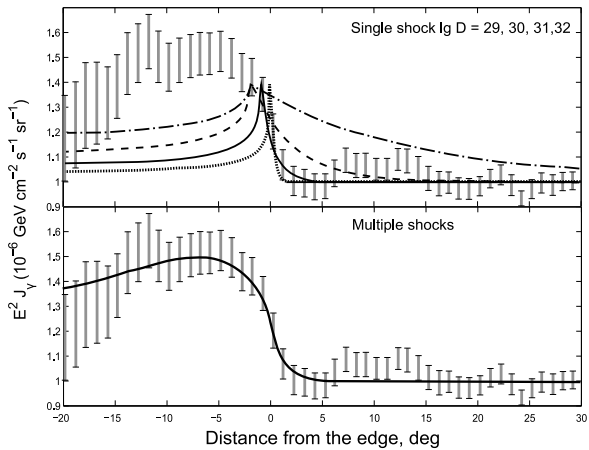} \caption{Spatial distribution of
the gamma-ray emission. Data are from \cite{meng}. {\it Top:} in case of single shock. Dotted line correspond to D=10$^{29}$ cm$^2$/s, solid to D=10$^{30}$ cm$^2$/s, dashed to D=10$^{31}$ cm$^2$/s and dash-dotted D=10$^{32}$ cm$^2$/s, {\it Bottom:} in case of several shocks distributed in accordance with (\ref{r_shock}).} \label{fig:spat}
\end{figure}
In Fig. \ref{fig:spat} we show an expected spatial distributions of
gamma-ray emission from the bubble for the
  single shock  and  multiple
shocks cases. For the multiple shocks it was assumed that they
distributed in accordance to the solution (\ref{r_shock}) and it
was also assumed that the amount of accelerated electrons is
proportional to the density of shocks. From this figure one can
see that the single shock model (without the second order Fermi acceleration
in the bubble interior) is unable to reproduce the  data. However,
for the parameters of star capture model these data are nicely
described.

\section{Conclusion}
We have shown that series of shocks produced by a sequential
stellar captures by the central black hole can further
re-accelerate the protons emitted by SNRs up to energies above
$10^{15} eV$.  The predicted CR spectrum contributed by the Bubble
may be $E^{-\nu}$ where $\nu \sim 3$ for $10^{15}\mbox{ eV} < E <
10^{19}\mbox{ eV}$ that explains the knee CR spectrum.

The regime of electron acceleration in the bubble is quite
different from that of protons. It is a combination of single and
multishock accelerations. In this case we have a cut-off of the
electron spectrum at $E> 3~10^{13}$ eV and flattening of the
spectrum at $E<100$ GeV that explains nicely the bubble gamma-ray
spectrum and the sharp edge spatial distribution observed by
\cite{meng} if this emission is due to IC on the relic photons
%\vspace{\baselineskip}

\section*{Acknowledgements}
 DOC and VAD are partly supported
by the NSC-RFBR Joint Research Project RP09N04 and 09-02-92000-HHC-a. KSC is
supported by the GRF Grants of the Government of the Hong Kong SAR under HKU
7011/10P. CMK is supported, in part, by the Taiwan National Science
Council Grant NSC 98-2923-M-008-01-MY3 and NSC 99-2112-M-008-015-MY3.
WHI is supported by the Taiwan National Science Council Grant NSC
97-2112-M-008-011-MY3 and Taiwan Ministry of Education under the Aim for
Top University Program National Central University.

\clearpage

\end{document}